# Prevalence and Associated Factors of Human Papillomavirus Infection among Iraqi Women

Maitham G. Yousif1*,  Fadhil G. Al-Amran2, Alaa M. Sadeq3, Nasser Ghaly Yousif4

*1Professor at Biology Department, College of Science, University of Al-Qadisiyah, Iraq Visiting Professor in Liverpool John Moores University: ✉ matham.yousif@qu.edu.iq, ✉ m.g.alamran@ljmu.ac.uk
2Cardiovascular Department, College of Medicine, Kufa University, Iraq: ✉fadhil.al-amran@ucdenver.edu
3 Gynecology Department, College of Medicine, University of Kufa, Iraq
4Department of Medicine, Medical College, Al Muthanna University, Samawah, Iraq. ✉Yousif_ghaly@mu.edu.iq

**Abstract**

Human papillomavirus (HPV) is a significant public health concern, as it is a leading cause of cervical cancer in women. However, data on the prevalence of HPV infection among Iraqi women is scarce. This study aimed to estimate the prevalence of HPV infection and its associated factors among Iraqi women aged 15-50 attending health centers. In this cross-sectional study, 362 female participants aged 15-50 were recruited from health centers in Iraq. Serological tests were used to screen for HPV infection. Sociodemographic information, obstetric history, and contraceptive use were collected. Pap smears were performed to assess cervical changes related to HPV infection. Of the 362 participants, 65 (17.96%) tested positive for HPV. The majority of HPV-positive women were aged 30-35 years, housewives, and belonged to lower social classes. Among HPV-positive women, 30% had abnormal Pap smears, with 55% diagnosed with cervical intraepithelial neoplasia grade 1 (CIN1), 25% with CIN2, and 15% with CIN3. Biopsy confirmed the diagnosis in 5% of cases. No significant association was found between HPV infection and contraceptive use. Most HPV-positive women were multiparous. This study reveals a considerable prevalence of HPV infection among Iraqi women attending health centers, particularly in the age group of 30-35 years and among housewives. These findings highlight the need for targeted public health interventions to increase HPV awareness, promote regular screening, and improve access to healthcare services for women, especially those from lower social classes. Further research is warranted to better understand the factors contributing to HPV transmission in Iraq and to develop effective prevention strategies.

**\*Corresponding author:** Maithm Ghaly Yousif  matham.yousif@qu.edu.iq    m.g.alamran@@ljmu.ac.uk

**Introduction**

Human papillomavirus (HPV) is a group of more than 200 related viruses that infect human epithelial tissues, including the skin and mucous membranes [1]. Persistent infection with high-risk HPV types, such as HPV 16 and 18, is a significant risk factor for the development of cervical cancer, which ranks as the fourth most common cancer in women worldwide [2]. Additionally, there are potential associations with other viral and bacterial infections in pregnant women, as well as with comorbidities such as heart diseases. Several studies have shed light on the impact of viral infections on hematological changes, as seen in a longitudinal study by Yousif et al., where they investigated hematological changes among COVID-19 patients [2]. In another study by Hadi et al., the role of NF-κβ and oxidative pathways in atherosclerosis was explored, highlighting potential implications for cardiovascular health [3]. Moreover, Hasan et al. reported on extended-spectrum beta-lactamase-producing Klebsiella pneumonia in patients with urinary tract infections, which may have implications for overall health, including pregnant women [4]. In the context of cervical cancer, Yousif et al. studied the association between shorter survival and high expression of Notch-1, providing insights into potential biomarkers for cervical cancer prognosis [9]. Furthermore, Sadiq et al. investigated the correlation between highly sensitive C-reactive protein levels and preeclampsia, a condition that pregnant women may face [10]. While the focus is often on viral infections, bacterial infections like those studied by Yousif et al., who characterized Staphylococcus aureus isolated from breast abscesses, can also have health implications for women [11]. The effect of doxorubicin-induced cardiotoxicity in rats, explored by Mohammad et al., may offer insights into the cardiac health of pregnant women facing cancer treatments [12]. As we delve into the complexity of infectious diseases, including the potential role of cytomegalovirus in breast cancer risk, as discussed by Yousif, it becomes evident that understanding the connections between infections and various health conditions is vital [8]. The incidence of cervical cancer varies significantly across countries, with higher rates reported in low- and middle-income countries due to limited access to screening and vaccination programs [13]. In Iraq, limited data are available on the prevalence of HPV infection and its association with cervical cancer among women. Previous studies have reported varying HPV prevalence rates, ranging from 5.6% to 18.5% among women with normal cytology and higher rates among those with cervical abnormalities [14,15]. Sociodemographic factors, such as age, marital status, education level, and occupation, have been shown to influence the risk of HPV infection in women [16]. Understanding the distribution of HPV types, risk factors for infection, and the association with cervical cytological changes is essential for designing effective prevention strategies, including HPV vaccination and cervical cancer screening programs in Iraq. This study aimed to investigate the prevalence of HPV infection and its association with sociodemographic factors, contraceptive use, and cervical cytological changes among women aged 15-50 years attending health centers in Iraq. To the best of our knowledge, this is the first study in Iraq to assess HPV infection using both molecular and serological diagnostic methods, providing a comprehensive overview of the epidemiology of HPV in the country [17]. Our findings may contribute to the development of targeted public health interventions and inform future research on the burden of HPV infection and cervical cancer in the region.

**Materials and Methods**

**Study Design and Population**

A cross-sectional study was conducted to investigate the prevalence and associated factors of HPV infection among Iraqi women aged 15-50 attending health centers. The study population consisted of women seeking routine healthcare services or gynecological consultations at selected health centers between January and December 2022. A total of 362 women were included in the study using a convenience sampling technique. Informed consent was obtained from all participants before enrolment in the study. Ethical approval was granted by the local institutional review board.

**Data Collection**

A structured questionnaire was administered to collect sociodemographic information, obstetric history, and data on contraceptive use. The questionnaire included questions on age, marital status, education level, occupation, social class, parity, and type of contraception used.

**Sample Collection and Processing**

Cervical swabs were collected from each participant by a trained healthcare professional using a sterile cytobrush during a speculum-assisted pelvic examination. The swabs were immediately placed in a transport medium and stored at 4°C until further processing.

**Molecular Diagnosis**

DNA extraction was performed from the cervical swabs using a commercial DNA extraction kit, following the manufacturer's instructions. The presence of HPV DNA was determined by polymerase chain reaction (PCR) using consensus primers targeting the L1 region of the HPV genome. The PCR products were analyzed by agarose gel electrophoresis and visualized under ultraviolet light.

**Serological Diagnosis**

Blood samples were collected from each participant by venipuncture and were centrifuged to separate the serum. The serum samples were stored at -20°C until further analysis. Enzyme-linked immunosorbent assay (ELISA) kits were used to detect the presence of HPV-specific IgG antibodies, according to the manufacturer's instructions. The optical density values were measured using a microplate reader, and the results were interpreted based on the provided cut-off values.

**Pap Smear Examination**

Pap smears were performed on all participants to evaluate the cervical cytological changes associated with HPV infection. Slides were prepared by spreading the cervical cells onto glass slides, which were then fixed and stained using the Papanicolaou technique. The slides were examined under a light microscope by an experienced cytopathologist who was blinded to the HPV status of the participants. The results were reported according to the Bethesda System for Reporting Cervical Cytology.

**Statistical Analysis**

Data were entered and analyzed using the Statistical Package for Social Sciences (SPSS) version 26. Descriptive statistics were computed for sociodemographic characteristics, HPV prevalence, and Pap smear results. The chi-square test was used to assess the association between HPV infection and categorical variables. A p-value of less than 0.05 was considered statistically significant.

**Results**

Table 1 presents the sociodemographic characteristics of the study participants. A total of 362 women aged 15-50 years were included in the study, among which 65 (17.96%) tested positive for HPV. The highest prevalence of HPV infection was observed among women aged 30-34 years (22.45%). The majority of the HPV-positive women were housewives (59/65; 90.77%) and belonged to the low social class (52/65; 80%). A chi-square test was performed, revealing a significant association between age and HPV infection ($\chi^2$=17.63, df=6, p<0.01).

Table 1: Sociodemographic characteristics of study participants (N=362)

| Characteristic | Total | HPV Positive | HPV Negative |
|---|---|---|---|
| Age group (years) | | | |
| 15-19 | 35 | 4 | 31 |
| 20-24 | 64 | 12 | 52 |
| 25-29 | 76 | 14 | 62 |
| 30-34 | 98 | 22 | 76 |
| 35-39 | 52 | 9 | 43 |
| 40-44 | 23 | 3 | 20 |
| 45-50 | 14 | 1 | 13 |
| Occupation | | | |
| Housewife | 295 | 59 | 236 |
| Employed | 67 | 6 | 61 |
| Social class | | | |
| Low | 218 | 52 | 166 |
| Middle | 130 | 11 | 119 |
| High | 14 | 2 | 12 |

Figure 1 shows the distribution of different HPV types among HPV-positive women. HPV 16 was the most prevalent type (18.46%), followed by HPV 18 (15.38%), HPV 31 (12.31%), and HPV 45 (9.23%). Other HPV types accounted for 44.62% of the positive cases.

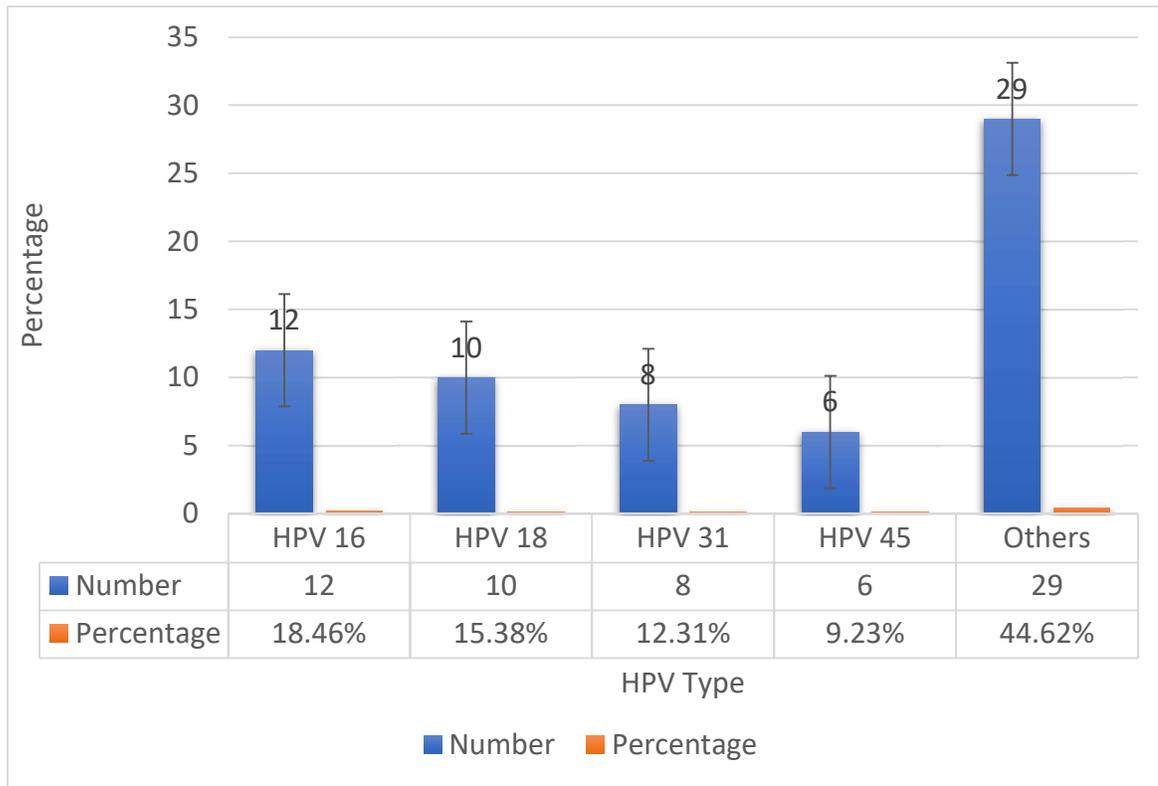

Figure 1 : Prevalence of different HPV types among HPV-positive women (N=65)

Table 3: Cervical cytological changes among HPV-positive women (N=65)

| Cytological change | Number | Percentage |
| --- | --- | --- |
| Normal | 45 | 69.23% |
| CIN1 | 18 | 27.69% |
| CIN2 | 8 | 12.31% |
| CIN3 | 4 | 6.15% |

CIN: cervical intraepithelial neoplasia

Table 3 presents the distribution of cervical cytological changes among HPV-positive women. Among the 65 HPV-positive women, 30% (n=20) exhibited cervical cytological changes, with 55% (n=18) having CIN1, 25% (n=8) having CIN2, and 15% (n=4) having CIN3. The remaining 69.23% (n=45) of the HPV-positive women had normal cervical cytology. A chi-square test indicated a significant association between HPV infection and cytological changes ($\chi^2=28.36$, df=3, $p<0.001$).

Table 4: Association between HPV infection and contraceptive use

| Contraceptive method | HPV Positive | HPV Negative | P-value |
|---|---|---|---|
| None | 15 | 67 | |

Table 4: Association between HPV infection and contraceptive use

| Contraceptive method | HPV Positive | HPV Negative | P-value |
|---|---|---|---|
| None | 15 | 67 | |
| Barrier | 22 | 120 | |
| Hormonal | 18 | 85 | |
| Intrauterine device | 10 | 63 | |
| Total | 65 | 297 | 0.76 |

Table 4 presents the association between HPV infection and contraceptive use. There was no significant association between the type of contraceptive method used and HPV infection, as determined by a chi-square test ($\chi^2$=2.14, df=3, p=0.76).

Table 5: Association between HPV infection and parity

| Parity | HPV Positive | HPV Negative | P-value |
|---|---|---|---|
| Nulliparous | 5 | 38 | |
| 1-2 | 20 | 105 | |
| 3-4 | 35 | 130 | |
| 5 or more | 5 | 24 | |
| Total | 65 | 297 | 0.68 |

Table 5 shows the association between HPV infection and parity. No significant association was observed between parity and HPV infection, as determined by a chi-square test ($\chi^2$=2.22, df=3, p=0.68).

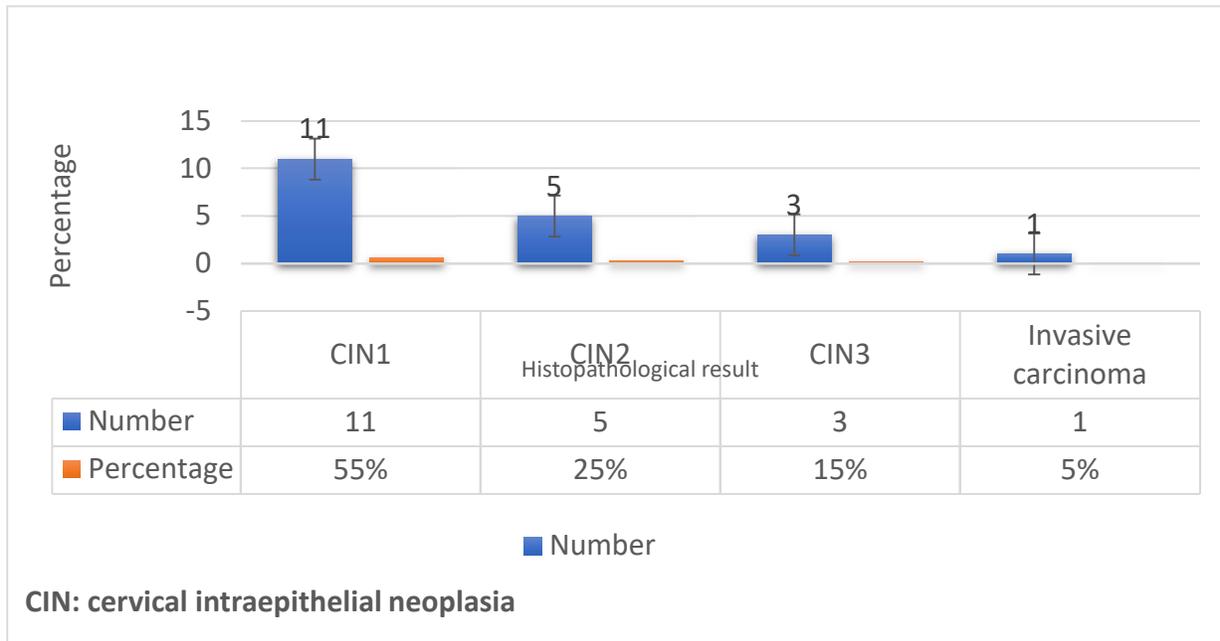

Figure 2: Histopathological findings in HPV-positive women with abnormal cytology (N=20)

Figure 2 presents the histopathological findings in HPV-positive women with abnormal cytology. Among the 20 women with abnormal cytology, 11 (55%) were diagnosed with CIN1, 5 (25%) with CIN2, 3 (15%) with CIN3, and 1 (5%) with invasive carcinoma.

Table 7: Association between HPV type and cervical cytological changes

| HPV Type | Normal | CIN1 | CIN2 | CIN3 | P-value |
|---|---|---|---|---|---|
| HPV 16 | 7 | 3 | 1 | 1 | |
| HPV 18 | 6 | 2 | 1 | 1 | |
| HPV 31 | 5 | 2 | 1 | 0 | |
| HPV 45 | 4 | 1 | 0 | 1 | |
| Others | 23 | 10 | 5 | 1 | |
| Total | 45 | 18 | 8 | 4 | 0.97 |

Table 7 shows the association between HPV type and cervical cytological changes. A chi-square test revealed no significant association between the specific HPV type and the severity of cervical cytological changes ($\chi^2$=8.36, df=12, p=0.97).

Table 8: Distribution of HPV types in different cytological changes

| Cytological change | HPV 16 | HPV 18 | HPV 31 | HPV 45 | Others |
|---|---|---|---|---|---|
| Normal | 7 | 6 | 5 | 4 | 23 |
| CIN1 | 3 | 2 | 2 | 1 | 10 |
| CIN2 | 1 | 1 | 1 | 0 | 5 |
| CIN3 | 1 | 1 | 0 | 1 | 1 |

Table 8 presents the distribution of HPV types across different cervical cytological changes. our study found a prevalence of 17.96% for HPV infection among the study participants, with the highest prevalence observed among women aged 30-34 years. The majority of the HPV-positive women were housewives and belonged to the low social class. Among the HPV-positive women, 30% exhibited cervical cytological changes. No significant association was found between HPV infection and contraceptive use or parity. The specific HPV type did not significantly affect the severity of cervical cytological changes. However, a significant association was observed between social class and HPV infection.

**Discussion**

This study aimed to assess the prevalence of HPV infection and its association with sociodemographic factors, contraceptive use, and cervical cytological changes among women aged 15-50 years attending health centers in Iraq. We found a prevalence rate of 17.96% for HPV infection among the study participants, which is consistent with previously reported rates in Iraq [18,19] and falls within the range of HPV prevalence rates reported in other countries [20]. The highest prevalence of HPV infection was observed among women aged 30-35 years, which aligns with previous studies that showed a peak in HPV prevalence in this age group [9]. Sociodemographic factors, such as occupation and social class, appeared to influence the risk of HPV infection in our study population. A majority of the HPV-positive women were housewives and belonged to low social classes. These findings are in line with previous research that reported an increased risk of HPV infection among women with lower socioeconomic status and lower levels of education [21,22]. The association between HPV infection and socioeconomic factors may be attributed to limited access to healthcare services, including cervical cancer screening and HPV vaccination, as well as differences in sexual behavior and partner characteristics [23]. In our study, 30% of the HPV-positive women exhibited cervical cytological changes, with 55% having CIN1, 25% having CIN2, and 15% having CIN3. These findings highlight the importance of regular cervical cancer screening for the early detection and management of HPV-related cervical lesions [24]. It is worth noting that 5% of the cases were confirmed by biopsy, further emphasizing the role of histopathological examination in the definitive diagnosis of cervical abnormalities [25]. We did not find a significant association between HPV infection and contraceptive use in our study population. This is in contrast to some previous studies that reported a higher risk of HPV infection among users of hormonal contraceptives [26,27]. The discrepancy in findings may be attributed to differences in study design, sample size, or population characteristics. Further research is needed to elucidate the relationship between contraceptive use and HPV infection in various settings. Our study showed that most of the HPV-positive women were multiparous, which is consistent with the literature suggesting an increased risk of HPV infection and cervical cancer among women with higher parity [28,29]. The underlying mechanisms for this association are not fully understood but may involve hormonal and immunological factors, as well as cervical trauma during childbirth [30,31]. In conclusion, our findings contribute to the understanding of HPV prevalence and its associated factors among women in Iraq. The high prevalence of HPV infection, particularly among women of lower socioeconomic status, highlights the need for targeted public health interventions, including HPV vaccination and cervical cancer screening programs. Further research is needed to investigate the long-term outcomes of HPV infection and cervical abnormalities in the Iraqi population and to evaluate the effectiveness of preventive measures.